\documentclass[useAMS,usenatbib,usegraphicx]{emulateapj}
\usepackage{times}
\usepackage{pifont}

\shorttitle{Catching early dust ejection in WISE~J1810--3305}
\shortauthors{Gandhi, Yamamura \& Takita}

\def\wise{{\em WISE}}
\def\akari{{\em AKARI}}

\def\spitzer{{\em Spitzer}}

\def\herschel{{\em Herschel}}
\def\hst{{\em HST}}
\def\iso{{\em ISO}}
\def\iras{{\em IRAS}}

\def\p{$\pm$}
\def\ltsim{\mathrel{\hbox{\rlap{\hbox{\lower4pt\hbox{$\sim$}}}\hbox{$<$}}}}
\def\gtsim{\mathrel{\hbox{\rlap{\hbox{\lower4pt\hbox{$\sim$}}}\hbox{$>$}}}}

\def\Msun{M$_{\odot}$}
\def\Lsun{L$_{\odot}$}
\def\micron{$\mu$m}
\def\araa{ARA\&A}
\def\aap{A\&A}

\def\mnras{MNRAS}
\def\apj{ApJ}

\def\aj{AJ}
\def\apjl{ApJL}

\def\pasj{PASJ}

\def\nh{$N_{\rm H}$}

\def\lbol{$L_{\rm Bol}$}

\def\gx339{GX~339--4}
\def\swiftj1753{SWIFT~J1753.5--0129}
\def\xtej1118{XTE~J1118+480}
\def\av{$R_{\rm V}$}
\def\av{$A_{\rm V}$}
\def\ebv{$E_{\rm B-V}$}

\def\wisej{WISE~J180956.27--330500.2}

\begin{document}
\slugcomment{ApJ Letters accepted for publication 2012 April 11 (Submitted 2012 March 13).}

\title{Dramatic infrared variability of WISE~J1810--3305: catching early-time dust ejection during the thermal pulse of an AGB star?}
\author{Poshak Gandhi, Issei Yamamura and Satoshi Takita}
\affil{Institute of Space and Astronautical Science, Japan Aerospace Exploration Agency, 3-1-1 Yoshinodai, chuo-ku, Sagamihara, Kanagawa 252-5210, Japan}

\label{firstpage}
\begin{abstract}
We present the discovery of a source with broadband infrared photometric characteristics similar to Sakurai's Object. \wisej\ (hereafter, J1810--3305) shows very red \wise\ colors, but a very blue 2MASS [$K$] vs. \wise\ [W1 (3.4 \micron)] color. It was not visible during the \iras\ era, but now has a 12~\micron\ flux well above the \iras\ point source catalog detection limit. There are also indications of variability in historical optical photographic plates, as well as in multi-epoch \akari\ mid-infrared measurements. The broadband infrared spectral energy distribution shape, post-\iras\ brightening and multiwavelength variability are all characteristics also shared by Sakurai's Object -- a post asymptotic giant branch (post-AGB) star which underwent a late thermal pulse and recently ejected massive envelopes of dust that are currently expanding and cooling. Optical progenitor colors suggest that J1810--3305 may have been of late spectral class. Its dramatic infrared brightening, and the detection of a late-type optical counterpart are consistent with a scenario in which we have caught an extremely massive dust ejection event (in 1998 or shortly before) during the thermal pulse of an AGB star, thus providing a unique opportunity to observe stellar evolution in this phase. J1810--3305 is the only source in the entire \wise\ preliminary data release with similar infrared SED and variability, emphasizing the rarity of such sources. Confirmation of its nature is of great importance. 
\end{abstract}

\section{Introduction}

A substantial fraction of cosmic dust is manufactured in late-type stars in the asymptotic giant branch (AGB) phase. Dust and gas are ejected into the interstellar medium at a rate depending upon stellar mass and varying along its evolutionary track (e.g., \citealt{vassiliadiswood93}). For moderate-mass stars ($\gtsim 3$\,M$_\odot$), mass-loss is thought to be quasi-continuous and to evolve gradually, whereas low-mass stars ($\sim 1$\,M$_\odot$) experience several explosive mass-loss events triggered by a Helium-burning \lq thermal pulse\rq\ (e.g., \citealt{schroeder98}). Remnants of such episodic shedding have been observed as detached circumstellar shells in radio observations (e.g., \citealt{olofsson00}, \citealt{yamamura93}) and the infrared \citep{kerschbaum10, izumiura11}. But catching ejection events at an early stage is rare given their brevity.

One source which has provided a \lq real-time\rq\ view of stellar evolution, dust formation and ejection is the so-called born-again post-AGB star, Sakurai's Object (hereafter, \lq Sakurai\rq; \citealt{duerbeck96}). In this class of objects, a white dwarf undergoes a Helium-burning flash resulting in a rejuvenation of the giant phase \citep{iben83_heliumflash}, manifesting as drastic changes in broadband spectral shape and intensity on timescales of days--months \citep{asplund99, hajduk05}. In the late nineties, Sakurai ejected a massive dust envelope which changed the source optical and infrared appearance completely, and which continues to evolve presently \citep{kerber99, kaufl03, evans06_sakurai, chesneau09}. 

Here, we present the discovery of similar massive dust ejection in a previously-unremarkable system, which appears to have been a late-type AGB star. This may be the first time that the early stage of a thermal pulse has been caught in this phase. The finding was made possible by combining new, wide-area infrared sky surveys. 

\section{Data}

\subsection{\wise}

The Wide-field Infrared Survey Explorer (\wise) satellite \citep[][]{wise} has carried out a highly-sensitive all-sky survey in four bands (W1--W4 centered on wavelengths of $\approx$3.4, 4.6, 12 and 22 \micron, respectively). The effective angular resolution corresponds to a Gaussian with full-width-at-half-maximum $\approx$6\arcsec\ in W1--W3, and 12\arcsec\ in W4. The whole sky was surveyed once between 2010 January and July. Data for about 57\%\ of the sky were publicly available as of 2012 February, and we used this preliminary data release\footnote{http://wise2.ipac.caltech.edu/docs/release/prelim/} in order to select sources. The release includes calibrated Vega source magnitudes. Sources brighter than mags of 8.0, 6.7, 3.8 and --0.4 in W1--4, respectively, begin to be affected by saturation. Profile-fitting is done using only unsaturated pixels, thus extending the dynamic range for point source brightness characterization into the saturated regime\footnote{http://wise2.ipac.caltech.edu/docs/release/prelim/expsup/sec4\_5c.html}. Magnitudes were converted to fluxes using standard zeropoints and color correction assuming a $B_{\nu}(283)$ SED relevant for the sources herein (Table 1 of \citealt{wise}). 

\subsection{Supporting data}
The Two Micron All Sky Survey (2MASS) mapped the sky in the near-IR $J$, $H$ and $K_s$ (hereafter, $K$) bands between 1997--2001 from ground-based telescopes at Mt. Hopkins, Arizona and Cerro Tololo, Chile \citep{2mass}. We used zeropoints from \citet{cohen03_2mass}.

The InfraRed Astronomy Satellite (\iras) carried out a pioneering all-sky survey at 12, 25, 60, and 100~\micron\ in 1983 \citep{iras}. A point source catalog (PSC) was produced which covered 96\% of the sky to approximate detection limits of 0.5, 0.5, 0.5 and 1.5 Jy, respectively, in the four bands. 

The Japanese mission \akari\ possessed two instruments: the InfraRed Camera (IRC; \citealt{akariirc}) and the Far-Infrared Surveyor (FIS; \citealt{akarifis}). Its main product was an all-sky catalog covering the IRC 9 and 18~\micron\ bands \citep{akari_allsky} and FIS 65, 90, 140, and 160~\micron\ bands \citep{akari_allsky_fis}, to 80\% completeness flux limits of 0.12 and 0.22 Jy (IRC) and 3.3, 0.43, 3.6, and 8.2 Jy (FIS), in the respective bands. Its survey lasted from 2006 May to 2007 August. We only used fluxes with a reported quality flag {\tt fqual}=3, indicating good photometric reliability. 

Optical photographic plate data and their digitized versions are available over the sky. These date over the $\sim$1970s--1990s. More details are given in \S~\ref{sec:optical}.

\section{Sample selection}
\label{sec:sampleselection}

During a study of sources with extremely-red \wise\ colors, we serendipitously found an intriguing source with red \wise\ colors but a very blue 2MASS $K$ vs. \wise\ W1 color, and no \iras\ counterpart. In order to investigate this further, we refined and applied the following selection criteria to the entire preliminary data release:

\begin{enumerate}
   \item $\exists$ a 2MASS ($K$) association within 3\arcsec\  of the \wise\ position.
   \item $K$--W1 $< -3.7$.
   \item W1--W2 $>3.0$.
   \item {\tt ph\_qual}=\lq AAAA\rq.
   \item $\nexists$ an \iras\ PSC counterpart within 1\arcmin.
\end{enumerate}

The first two criteria select sources with an apparent 2MASS $K$-band flux (in $\nu F_\nu$ units) higher than that in \wise\ W1 by at least 100, either due to an intrinsically blue spectral slope, or due to variability between the 2MASS and WISE observation epochs. 

The third criterion selects sources with \wise\ colors characteristic of a cold or reddened SED. This disfavors sources with an intrinsically blue slope. 

{\tt ph\_qual} in the fourth criterion is a quality flag assuring secure photometry in all \wise\ bands (signal-to-noise S/N$>$10), so that we may explore SED characteristics in detail. 

The last criterion favors newly-brightened sources or those without published identifications. 

\section{Results}

\begin{figure*}
  \begin{center}
  \includegraphics[width=8.28cm,angle=0]{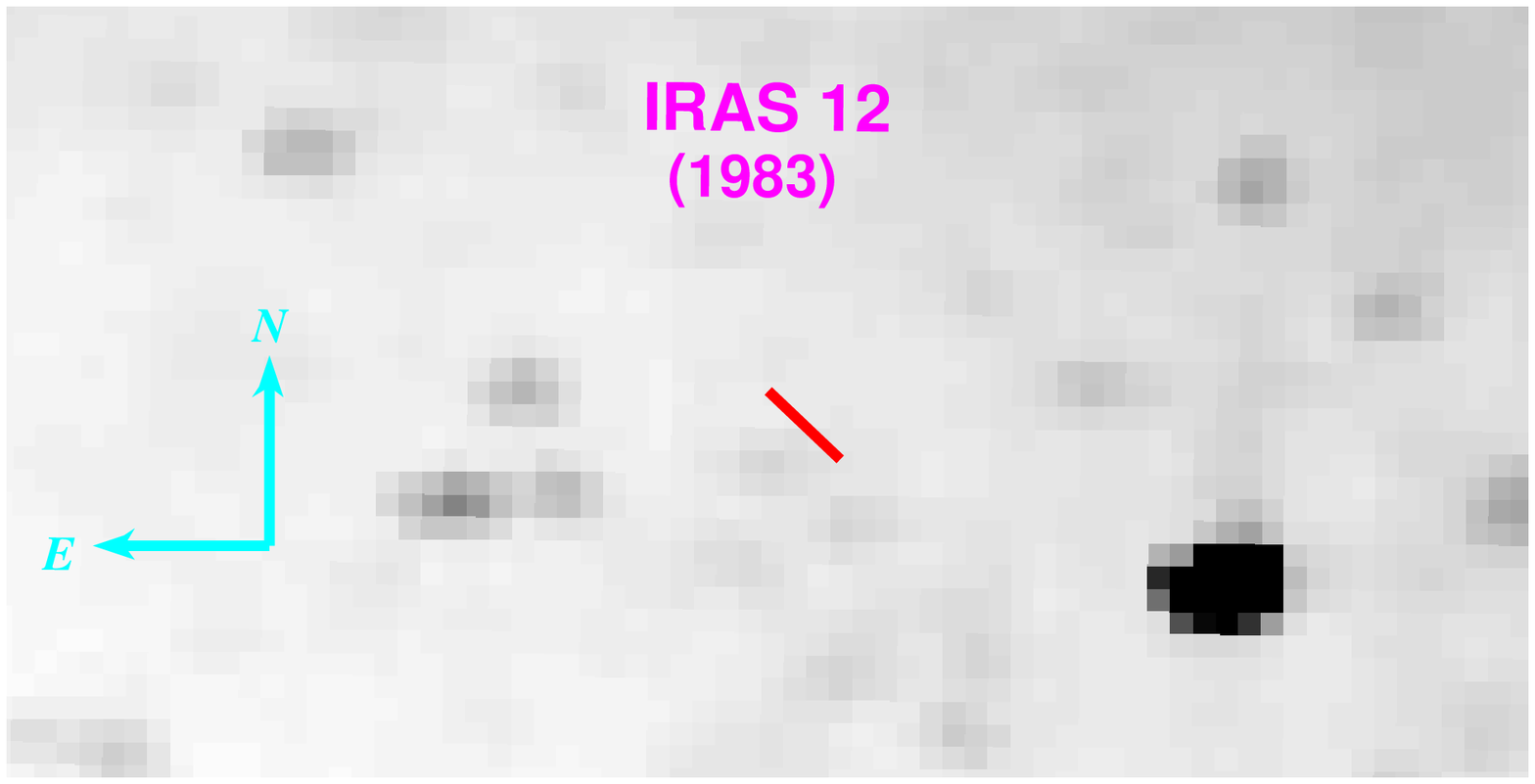}
  \includegraphics[width=16.8cm,angle=0]{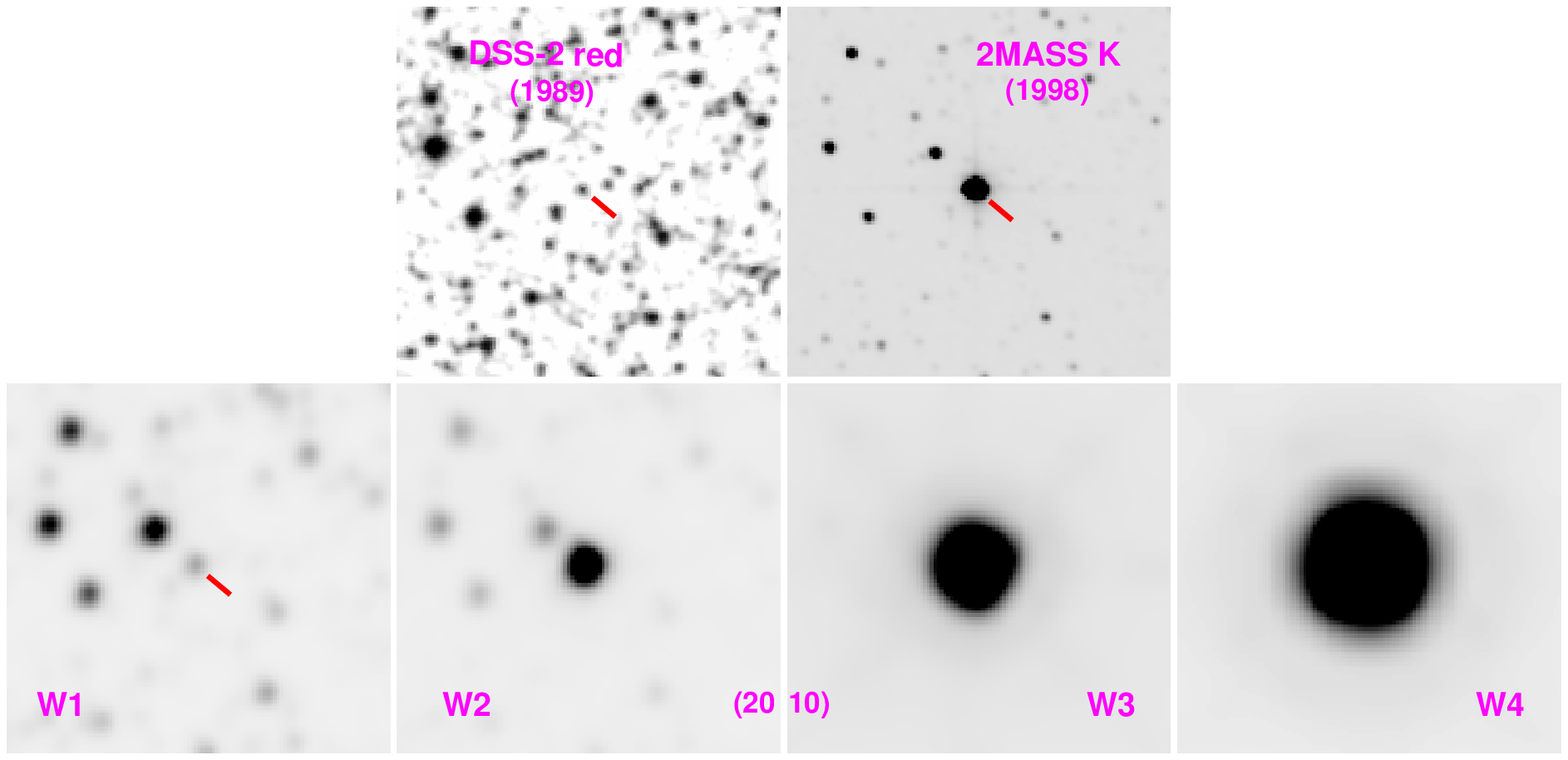}
  \caption{\iras\ (12~\micron), DSS2 (red), 2MASS ($K$) and \wise\ (W1--W4) images of the field of WISE J1810--3305. All images are centered on the source position (marked by a red dash in some). The \iras\ image is 100\arcmin$\times$50\arcmin\ wide; the rest are 2\arcmin .5 wide each. 
    \label{fig:images}}
  \end{center}
\end{figure*}

Of $\sim$257 million objects in the preliminary data release, only 38 satisfy the \wise+2MASS criteria 1--4 above. These are mostly stars of late (KM) spectral type along with a handful of Mira Cet stars and Carbon stars. Examining our final selection criterion, i.e. the presence of an \iras\ PSC counterpart, it turns out that two sources stand apart from the others. These are Sakurai and a source designated as \wisej\ (hereafter, \lq J1810--3305\rq). These are the only two sources {\em without} an \iras\ PSC counterpart. A comparison of the \wise\ and \iras\ fluxes is plotted in Fig.~\ref{fig:colmag} (interestingly, these two sources also stand out in terms of their red W3--W4 color, as shown). Assuming a conservative PSC flux limit of $F_{IRAS}^{\rm limit}$=0.5 Jy and neglecting small color corrections in the \iras\ 12~\micron\ vs. \wise\ W3 bandpasses, the above non-detections imply a brightening in the post-\iras\ 12~\micron\ fluxes by factors of at least 50 (Sakurai) and 13 (J1810--3305), respectively. The W3 fluxes of both lie well above the \iras\ PSC limit, and neither is confused with other objects of similar magnitude within a few arcmin, confirmed by careful inspection of the colors and fluxes of neighboring sources. Thumbnail images for J1810--3305 are shown in Fig.~\ref{fig:images}.

The infrared SEDs of both objects are plotted in Fig.~\ref{fig:sed}, including also \akari\ data. Photometry in the various bands is listed in Table~\ref{tab:photometry}. The deep \lq dip\rq\ at W1 in the SEDs of both sources (according to the selection of mismatched 2MASS and \wise\ colors) is apparent. 

\subsection{Optical counterpart}
\label{sec:optical}

\begin{figure*}
  \begin{center}
  \includegraphics[width=8.5cm,angle=90]{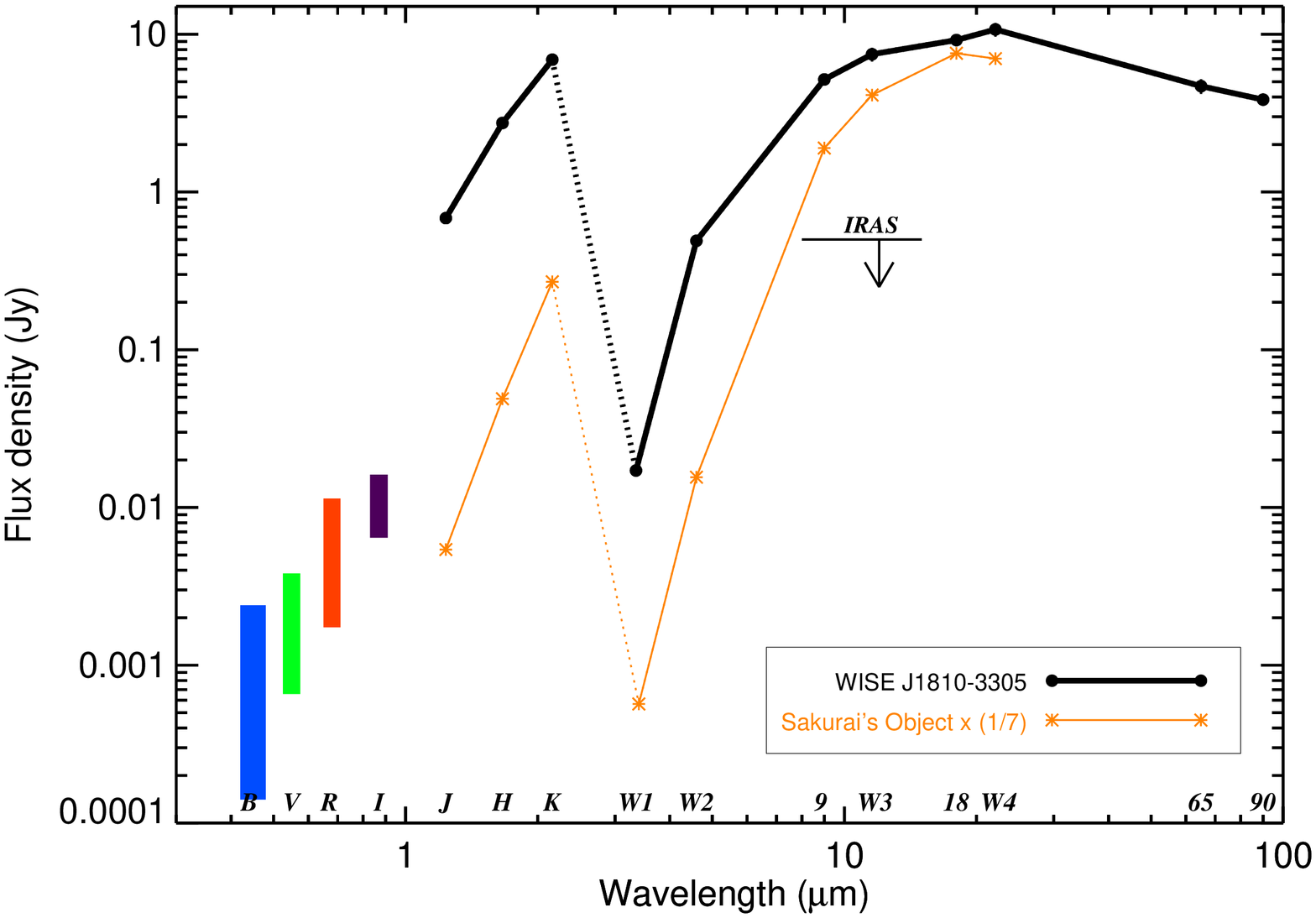}
  \includegraphics[width=8.5cm,angle=90]{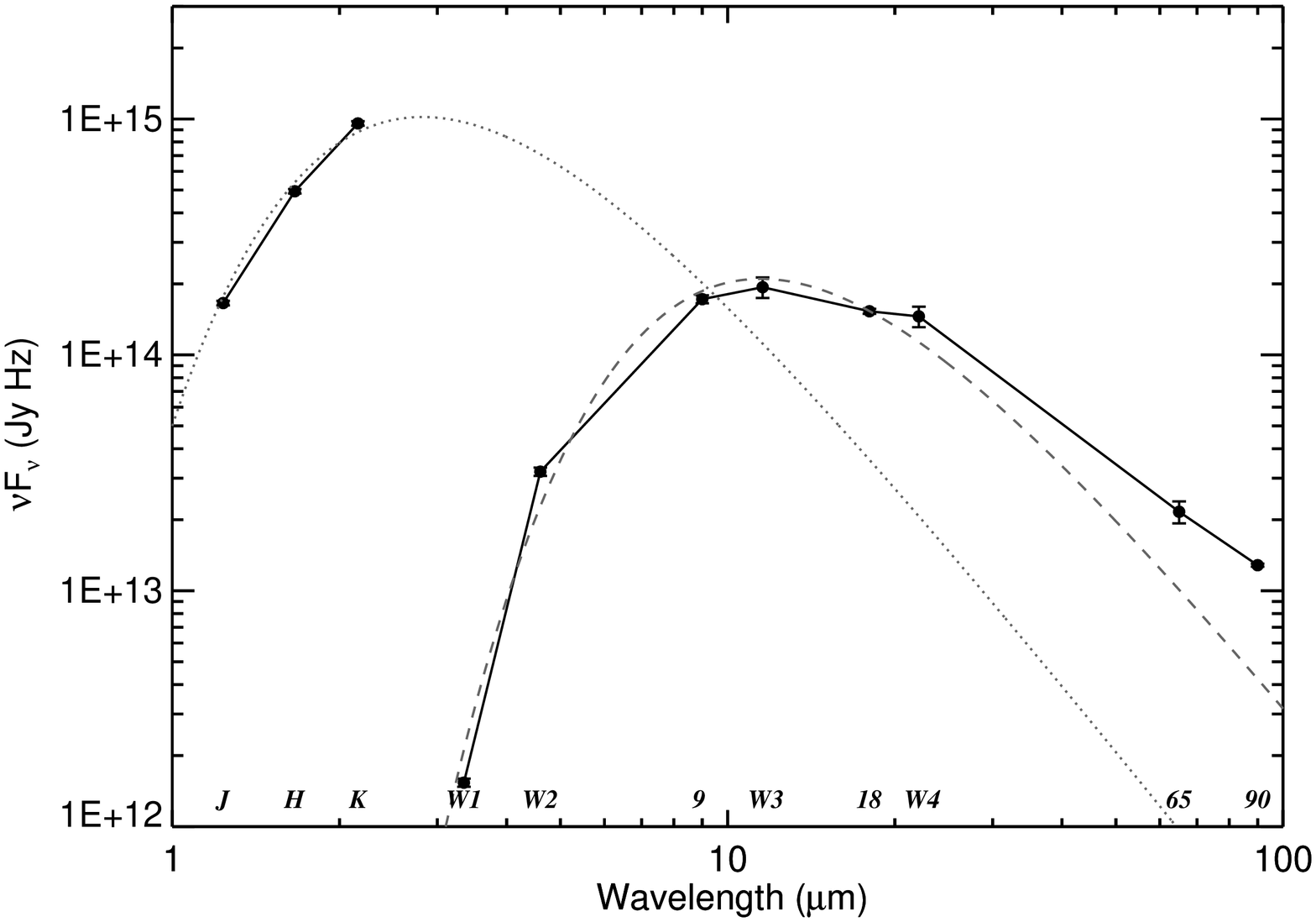}
  \caption{{\textbf{\textsl{(Top)}}} Broadband infrared SED of WISE~J1810--3305 (top thick lines). Dotted lines separate older 2MASS data from newer mid-IR (\wise+\akari) observations. Optical data are denoted by the color bars at the bottom left, and include \p0.5 mags to account for variability and systematic photometric calibration uncertainty. The SED of Sakurai's Object, scaled down by a factor of 7 for clarity, is shown as the lower thin (orange) lines. {\textbf{\textsl{(Bottom)}}} Planck functions overlaid on the IR SED of J1810--3305 in $\nu F_\nu$ units. The dotted curve covers the 2MASS points and has a temperature ($T$) of 1300~K, whereas the dashed curve over the \wise\ data is for $T$=320 K. These models are for illustrative purposes only, in order to emphasize the overall change in $T$, and the \akari\ far-IR excess. 
    \label{fig:sed}}
  \end{center}
\end{figure*}

Table~\ref{tab:photometry} lists magnitudes of the optical counterpart of J1810--3305 drawn from historical photographic surveys in $BVRI$. These must be used with caution, given that faint object measurement on old plates is subject to increased uncertainty as compared to CCDs. We preferred modern digitized scans where available\footnote{This was true in the $R$ band, where an older USNO--B1.0 measurement is $R2$=15.98 \citep{usnob}, about 2 mags fainter than in the digitized sky survey (DSS). We measured source flux in the DSS image and compared it to to field stars, obtaining consistent results with the brighter mag listed in Table~\ref{tab:photometry}.}. Furthermore, the exact epoch of observation is not always explicitly stated in the publicly available data. Magnitudes were converted to approximate fluxes using generic zeropoints\footnote{http://ssc.spitzer.caltech.edu/warmmission/propkit/pet/magtojy/} for plotting in Fig.~\ref{fig:sed}. Exact bandpass responses are not always available, so these should be considered approximate only.

\subsection{Multiwavelength variability}
\label{sec:variability}

Despite the above caveats, there are indications of strong multiwavelength optical variability. J1810--3305 appears to have brightened by $\sim$2 mags over the approximate period of 1968--1984 (exact dates unknown) in $B$ and $\sim$1 mag over the 1980s in $R$. The two available photometric data points in $V$ instead imply fading by $\sim$1 mag over $\sim$20 years preceding 1987.  

\akari\ observed the source over two epochs. The 9 and 18~\micron\ fluxes of J1810--3305 declined from 5.6 to 5.0 Jy and from 9.6 to 9.0 Jy, respectively, between 2006 September and 2007 March. Typical detection S/N$\ge$750, so these changes are significant.

The source position does not appear to have been covered\footnote{Referee communication} with other missions, e.g. \spitzer, \iso, \herschel, or \hst. 

\begin{figure}
  \begin{center}
  \includegraphics[width=6.5cm,angle=90]{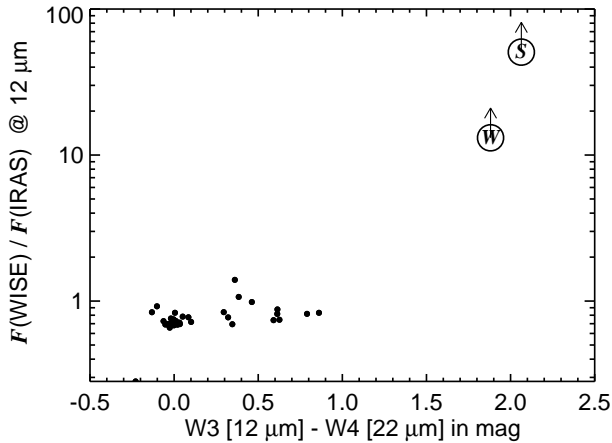}
  \caption{
Ratio of \wise\ to \iras\ 12~\micron\ flux densities vs. \wise\ color (W3--W4) 
for the 38 sources that satisfy our selection criteria 1--4. 
J1810--3305 (W) and Sakurai (S) stand out as having brightened dramatically at 12~\micron, and also in terms of their red W3--W4 color. 
    \label{fig:colmag}}
  \end{center}
\end{figure}

\section{Discussion}
\label{sec:discussion}

\subsection{Comparison of J1810--3305 and Sakurai} 
\label{sec:comparison}

The similarity in the broadband infrared properties and variability of J1810--3305 and Sakurai is striking. In particular, the 2MASS+\wise\ SEDs appear to mirror each other well, including the red \wise\ colors and the apparent W1 dip. Furthermore, both sources have only recently emerged in the mid-IR. As illustrated by Fig.~\ref{fig:colmag}, these properties are unique to J1810--3305 and Sakurai. The nature of J1810--3305 is completely unknown, but these facts suggest that its IR photometric properties are analogous to Sakurai. 

Sakurai is known to have ejected a large amount of dust around 1998 \citep[e.g. ][]{geballe02}. If J1810--3305 also ejected dust recently, the optical and near-IR are expected to have faded and reddened since peak brightness, while the mid and far-IR fluxes would rise due to dust reemission. The extremely red W3--W4 color of both these sources in Fig.~\ref{fig:colmag} is indicative of the presence of dust. 

The bright near-IR 2MASS fluxes are consistent with hot ($T$$\sim$1300 K) dust having been freshly ejected in or before 1998 (Fig.~\ref{fig:sed}). Such dust would have been visible with \iras\ at 12~\micron\ for any modified blackbody with emissivity beta-index $\beta$$<$2.5 ($F_\nu \propto \nu^\beta\ B_{\nu,T}$, where $F_\nu$ is observed flux density and $B_{\nu,T}$ is the Planck function). This suggests that dust ejection occurred sometime between the \iras\ and 2MASS epochs.

The \wise\ SED implies a much lower thermal temperature ($T$$\sim$320 K) in 2010, though a single modified blackbody SED does not fit the observed fluxes well (a mix of multi-temperature components is likely). The apparent 2MASS/\wise\ \lq mismatch\rq\ then implies some dust cooling over the $\sim$11+ interim years.

Integrating over the two blackbody curves in Fig.~\ref{fig:sed} gives bolometric infrared luminosities (\lbol) of $\sim$425$D^2$ \Lsun\ for the prior 2MASS component and $\sim$90$D^2$ \Lsun\ for the recent \wise+\akari\ component, where $D$ is the distance in kpc. Unlike Sakurai, where \lbol\ has stayed approximately constant, the bolometric IR power of J1810--3305 has declined by a factor of $\sim$5 between the 2MASS and \wise\ epochs. 

Regarding other wavelengths, in the optical a historical counterpart exists for J1810--3305, but was not known pre-1991 for Sakurai. Multiwavelength variability is a common characteristic in both objects. In the far-IR regime, the \akari\ 65 and 90~\micron\ data of J1810--3305 lie above the extrapolation of the blackbody curve overlaid on the \wise\ data (see Fig.~\ref{fig:sed}), suggesting the presence of additional cooler dust already in existence in 2006, or some systematic change in dust properties. Unfortunately, \akari\ did not carry out sufficient scans of Sakurai for accurate far-IR photometry.

\subsection{Extinction}
\label{sec:extinction}

J1810--3305 lies more than 6 degrees below the Galactic plane ($l$=--1$^\circ$.0, $b$=--6$^\circ$.6). The integrated Galactic neutral gas column (\nh) along this line-of-sight is determined to be \nh=1.88$\times$10$^{21}$ cm$^{-2}$ \citep{dickeylongman90}. For a typical Milky Way dust:gas ratio, this translates into a reddening value of \ebv=0.32 \citep{bohlin78}, or an optical extinction \av=1.0 for a standard extinction curve slope with $R_{\rm V}$=3.1. Near- and mid-IR extinction is 10--20 times smaller still (see, e.g., several references listed in \S~2.2.3 of \citealt{g11_wise}). The actual extinction could either be lower if some of the gas column lies behind J1810--3305, or may be somewhat higher if $R_{\rm V}$ happens to be larger along this line-of-sight. In any case, these numbers suggest that interstellar reddening is unlikely to be extreme, so the SEDs plotted in Fig.~\ref{fig:sed} should not change dramatically after extinction correction, especially in the IR.

\subsection{Dust mass}
\label{sec:dustmass}

Dust mass may be computed by using the infrared emissivity relation of \citet{hildebrand83}, given $D$ and the dust characteristics. Here, we simply assume a canonical graphite composition with 0.1~\micron-sized grains (Carbon-rich grains have been identified around Sakurai; \citealt[e.g. ][]{eyres98, evans06_sakurai}). Specifically, we use Eq.~2 from \citet{g11_m82}. For the recent \wise\ data with $T$$\sim$320~K, the observed W3 flux is $F_{12~\mu \rm m}$=7.5 Jy. This yields a dust mass of $M_{\rm d}$=2.1$\times$10$^{-7}$$D^2$ \Msun. If $D$=2--5 kpc similar to Sakurai \citep{vanhoof07}, $M_{\rm d}$=(0.9--5)$\times$10$^{-6}$ \Msun\ for J1810--3305. For a typical gas:dust mass ratio of $\gtsim$100, the implied total mass-loss rate is large \citep[cf. ][]{yamamura93,winters00}, suggesting it cannot remain steady. A non-zero $\beta$ index would boost the mass further. 
 
These numbers must be used with caution since the absolute values of dust physical parameters are presently unknown. Instead, a comparison with Sakurai may be useful, for which we find $M_{\rm Sakurai}(\rm 2010\ dust)$$\approx$3$\times$10$^{-6}$$D_{\rm Sakurai}^2$ \Msun. Thus, $M_{\rm J1810-33}/M_{\rm Sakurai}$=0.07$(D_{\rm J1810-33}/D_{\rm Sakurai})^2$. 

\subsection{The nature of J1810--3305}
\label{sec:nature}

A major uncertainty is the spectral type of the central star. Assuming a reddening of \ebv=0.3, $B$--$V$ varies from $\approx$2.4 to --0.6, $V$--$R$ is about 1--2 and $R$--$I$$\approx$0.5 or 1.5, where color measurements are based upon photometry closest matched in time. Uncertainties are $\approx$0.7--1 mag. Photometric variability and other caveats mentioned in \S~\ref{sec:optical} and \S~\ref{sec:variability} complicate the issue. Fig.~\ref{fig:sed} plots the plausible range of observed optical fluxes after accounting for these uncertainties. The overall SED shape in the optical appears to have been red, and the redder colors suggest a tentative late (K or M) type classification. J1810--3305 may thus be in an earlier evolutionary phase as compared to the post-AGB phase that Sakurai was discovered in. 

Sakurai displayed a sharp increase in its optical fluxes around 1991 \citep{duerbeck96}. The rising optical phase for J1810--3305 may have already begun before the mid-1980s, according to its brightening $B$ and $R$ magnitudes. If the historical optical data probe pre-dust-ejection photospheric fluxes of a late-type central star with an SED peak around $\sim$1~\micron, then \lbol\ should have brightened by $\sim$100 in order to power the hot 2MASS emission (Fig.~\ref{fig:sed}). 

The emerging picture is that we could be observing early dust ejection during the thermal pulse of an AGB star. The presence of an optical pre-cursor, its (tentative) late-type classification, the rapid and strong increase in \lbol\ and the massive amount of dust ejected ($M_{\rm d}$$\sim$10$^{-6}$ \Msun) are all consistent with this scenario. The dust ejection epoch ($t_0$) lies between the \iras\ (1983) and 2MASS (1998) observations. 

$t_0$ may be estimated by assuming that the dust exists in a single shell in instantaneous thermal equilibrium with stellar heating. Based upon the approximate \lbol\ and $T$ values derived in \S~\ref{sec:comparison}, the shell radius $r$$\propto$\lbol$^{0.5}$$T^{-2}$, implying $r_{\rm ({\it WISE})}$=7.6$\times$$r_{\rm (2MASS)}$. This expansion took 11.7 years. If the expansion velocity remained constant since $t_0$, backward extrapolation places $t_0$ between 1996 October and 1998 July. Using a more realistic $\beta$ increases $r_{\rm({\it WISE})}/r_{\rm (2MASS)}$ and pushes $t_0$ closer to the 2MASS epoch. 

The $V$ band, on the other hand, shows a mild historical decline. This may be a sign of erratic (early-time) photometric variability such as Sakurai exhibited at late times \citep{duerbeck00}, though the large photometric uncertainties make conclusive inferences impossible. If \lbol\ did increase $\sim$100-fold post-1983, it would imply a peak $V$ mag $\sim$10.5--11.4, similar to Sakurai. If so, why was the appearance of a new bright star not reported for J1810--3305? One possibility is a very brief period between brightening and dust obscuration. Moreover, the source location is far enough outside the Galactic plane that it does not fall in intensively-monitored fields, but is still close enough to the plane that the field is relatively crowded in the optical making it a non-trivial study for amateur observations. 

The rapid evolution of Sakurai outpaces all canonical predictions made by theory of post-AGB stars, which expects slower flux changes from the onset of the Helium flash \citep{iben83_heliumflash, kerber99_proc, herwig01}. In contrast, J1810--3305 may now be providing us a testbed for studying the Helium flash of the (earlier) AGB phase. The infrared brightening and reddening appear to be coeval in both objects, but the earlier optical rise of J1810--3305 suggests that the optical evolution of J1810--3305 may be slightly slower than that of Sakurai (\S~\ref{sec:optical}). Furthermore, the post-2MASS \lbol\ decline indicates that the central star could already be relaxing following the thermal pulse, unlike the apparently steady output of Sakurai. If so, the thermal pulse lasted no more than $\sim$30 years -- the interval from the optical measurements of the late-1980s to the \wise\ epoch. This is exceptionally-brief even compared to the \lq ultra-short\rq\ mass-loss timescales of $\sim$10$^2$ years considered in the literature (e.g. \citealt{iben82,schroeder98}). The other possibility is that multi-temperature expanding shells result in complex optical depth effects unaccounted for in the \lbol\ computation.

Large sky surveys are optimal for serendipitously catching rare objects. The fact that only two sources (J1810--3305 and Sakurai) in the entire \wise\ preliminary database share similar properties emphasizes the rare (brief) nature of such events. Only a handful of other sources have been compared to Sakurai (e.g. A~30, A~78, FG Sge, V605 Aql and V838 Mon; \citealt{jacobyford83, lawlor05, guerrero12}). Almost all are in more advanced states of evolution, and none appears to be coeval with Sakurai. J1810--3305 thus provides a unique opportunity for follow-up. 

Ascertaining the degree of affinity between these sources requires optical and infrared spectroscopic observations of J1810--3305 in order to determine its spectral class, measure element abundances and understand the nature of circumstellar dust. Irrespective of these comparisons, however, it appears that we have identified a massive dust ejection event at early times, making WISE J1810--3305 an interesting object in its own right.

\vspace*{1cm}
\acknowledgements
\noindent
PG acknowledges a JAXA International Top Young Fellowship, and ST a JSPS Postdoctoral Fellowship. \wise\ is a project of Univ. California, Los Angeles, and Jet Propulsion Laboratory (JPL)/California Institute of Technology (Caltech), funded by the National Aeronautics and Space Administration (NASA). 2MASS is a project of the Univ. Massachusetts and the Infrared Processing and Analysis Center/Caltech, funded by NASA and the National Science Foundation. \akari\ is a project of the Japan Aerospace Exploration Agency with the participation of the European Space Agency. Data products from the \iras\ missions were important for the analysis herein. The NASA/IPAC Infrared Science Archive (IRSA) operated by JPL under contract with NASA, and SIMBAD operated at CDS, Strasbourg, France, were the main databases queried. The USNO Image and Catalogue Archive is operated by the U.S. Naval Observatory. The DSS was produced at the Space Telescope Science Institute under U.S. Government grant NAG W-2166, and the UK Schmidt Telescope operated by the Royal Observatory Edinburgh and by the Anglo-Australian Observatory. The expert referee's review was prompt and valuable.

\begin{table*}[h]
  \begin{center}
    \caption{Photometry of WISE J1810--3305 and Sakurai's Object\label{tab:photometry}}
{
  \begin{tabular}{l|ccc|ccr}
    \hline
              & \multicolumn{3}{c}{{\underline{J1810--3305}}}&        \multicolumn{3}{c}{{\underline{Sakurai's Object}}}\\
    Band      &   Epoch          &  Mag       &   Flux density &    Epoch    &  Mag  & Flux density\\
              &                  &  Vega      &       Jy       &             &  Vega &       Jy\\       
    \hline
    B$^{[1]}$  &   1966--74 (1968?)& 18.17\p0.48 &                &             &        &        \\      
    B2$^{[2]}$ &   1978--90 (1984?)& 16.09       &                &             &        &        \\      
    V$^{[1]}$  &   1966--74 (1968?)& 15.49       &                &             &        &        \\      
    V$^{[3]}$  &   1987.710  & 16.40\p0.47 &                &             &        &        \\      
    R1$^{[2]}$ &   1974--87 (1984.2?)& 15.07       &                &             &        &        \\      
    R$^{[3]}$  &   1989.677    & 14.03\p0.45 &                &             &        &        \\      
    I$^{[2]}$  &   1978--2002 (1990?) & 13.36\p0.46 &                &             &        &        \\      
    2MASS J   &    1998.57        & 8.42\p0.02  &     0.68\p0.02 &   1999.33   &  11.56\p0.02 &  0.037\p0.001      \\
    2MASS H   &    1998.57        & 6.43\p0.02  &     2.74\p0.06 &   1999.33   &   8.69\p0.03 &  0.34\p0.01      \\
    2MASS K   &    1998.57        & 4.96\p0.02  &     6.90\p0.15 &   1999.33   &   6.37\p0.02 &  1.84\p0.04      \\
    AKARI 9   &    2006/09--2007/03 (2006.92)&  &     5.17\p0.20 &   2006/09--2007/03 (2006.97)   &        & 13.26\p0.65\\
    AKARI 18  &    2006/09--2007/03 (2006.92)&  &     9.19\p0.23 &   2006/09--2007/03 (2006.97)   &        & 53.01\p0.81\\
    AKARI 65  &    2006/09--2007/03 (2006.92)&  &     4.69\p0.50 &    ...      &  ...   & ...\\
    AKARI 90  &    2006/09--2007/03 (2006.92)&  &     3.85\p0.06 &    ...      &  ...   & ... \\
    W1        &    2010/03 (2010.222)& 10.27\p0.03 &     0.017\p0.001& 2010/03 (2010.215)&11.86\p0.07 & (4.0\p0.3)$\times$10$^{-3}$ \\
    W2        &    2010/03 (2010.222)& 6.24\p0.02  &     0.49\p0.01 &  2010/03 (2010.215)   &  7.87\p0.02  & 0.109\p0.002 \\
    W3        &    2010/03 (2010.222)& 1.62\p0.01  &     7.47\p0.07 &  2010/03 (2010.215)   &  0.15\p0.003 & 28.8\p0.1       \\
    W4        &    2010/03 (2010.222)& --0.27\p0.01&     10.7\p0.1  &  2010/03 (2010.215)   &  --1.92\p0.03& 49.0\p1.3       \\
    \hline
  \end{tabular}
}
~\par
  $^{[1]}$NOMAD/YB6/SPM (Bandpasses 103aO and 103aG+OG515 for $B$ and $V$); \citet{zacharias05}, \citet{girard98}; $^{[2]}$USNO-B1.0: (Surveys: $R1$: ESO--R; $B2$: SERC-J; $I$: SERC-I); \citet{usnob}; $^{[3]}$DSS+GSC2.3: (XV/V495 and AAO-SES/XS/IIIa-F+OG590 for $V$ and $R$); \citet{gsc}. For multiple observations (e.g. \akari/\wise), the median epoch is stated in brackets, and flux is the mean reported value. For historical optical surveys, the exact observation epoch is not always available; uncertain epochs or the median date of the full survey are stated with \lq ?\rq. For the DSS, the {\sc date-obs} fits header keyword is stated. Quoted photometric errors, where available, are statistical. 
  \end{center}
\end{table*}

\bibliographystyle{apj}

\label{lastpage}
\end{document}